\documentclass[preprin,showpacs,preprintnumbers,eqsecnum,amsmath,amssymb,nofootinbib]{revtex4}
\usepackage{graphics,epsfig,xcolor}
\usepackage{graphicx}
\usepackage{dcolumn}
\usepackage{bm}

\begin{document}
\baselineskip=0.8 cm
\title{High dimensional AdS-like black hole and phase transition in Einstein-bumblebee gravity}

\author{Chikun Ding$^{1,2,3}$}\thanks{Corresponding author, email: dingchikun@163.com;  dck@hhtc.edu.cn}
\author{Yu Shi$^{1}$}\author{Jun Chen$^{1}$}\author{Yuebing Zhou$^{1}$}
\author{Changqing Liu$^{1}$}
\affiliation{$^1$Department of Physics, Huaihua University, Huaihua, 418008, P. R. China\\
$^2$Department of Physics, Hunan University of Humanities, Science and Technology, Loudi, Hunan
417000, P. R. China\\
$^3$Key Laboratory of Low Dimensional
Quantum Structures and Quantum Control of Ministry of Education,
and Synergetic Innovation Center for Quantum Effects and Applications,
Hunan Normal University, Changsha, Hunan 410081, P. R. China}

\vspace*{0.2cm}
\begin{abstract}
\baselineskip=0.6 cm
\begin{center}
{\bf Abstract}
\end{center}

In this paper we obtain an exact high dimensional anti-de Sitter (AdS) black hole solution in Einstein-bumblebee gravity theory. This AdS-like black hole can only exist with a linear functional  potential of the bumblebee field. We find that the Smarr formula and the first law of black hole thermodynamics can still be constructed in this Lorentz symmetry breaking black hole spacetime as long as its temperature, entropy and volume are slightly modified. We find also that there exist two kinds of phase transition: small-large black hole phase transition and Hawking-Page phase transition, like those of Schwarzschild AdS black hole. After Lorentz symmetry breaking, the black hole mass at divergent point of heat capacity becomes small, and the Gibbs free energy of the meta-stable large black hole is also smaller, showing that the large stable black hole can be more easily formed.

\end{abstract}

 \maketitle

\vspace*{0.2cm}
\section{Introduction}
Local Lorentz invariance (LI) is the most fundamental spacetime symmetry of the special/general relativity (SR/GR) and the standard model (SM) of particle physics \cite{ashtekar}. LI means that the equations of motion for particles and fields are invariant under all Lorentz transformations(global for SR, local for GR) and diffeomorphism transformation for a particle. However, LI \textcolor[rgb]{0.00,0.00,1.00}{may fail to be} an exact principle at very high energy scales \cite{mattingly}. GR and SM are the best theories describing the four fundamental forces and there are no known experimental conflicts. Nonetheless, they are fundamentally different in that SM is a quantum theory  disregarding the gravitational effects of all particles, while GR is a classical geometrical theory ignoring all the quantum features of particles. For particles with energies on the order of $10^{30}$ eV (above the Planck scale), the gravitational interactions estimated by GR are very powerful, and so one shouldn't ignore it \cite{camelia}. Thus, on the very high energy scales, one has to reconsider combining the SM with GR in a unified theory, i.e., "quantum gravity".  An effective field theory combining GR and SM on low energy scales is called standard model extension (SME), which couples the SM to GR, and involves extra items embracing information about the violations of Lorentz invariance(Lorentz violation, LV) happening on the Planck scale \cite{kostelecky2004}.

Studying the Lorentz violation (LV) is a useful approach toward investigating the foundations of modern physics. Besides SME, string theory \cite{kostelecky198939}, noncommutative field theories \cite{carroll}, spacetime-varying fields \cite{bertolami69}, loop quantum gravity theory \cite{gambini}, brane world scenarios \cite{burgess03}, massive gravity \cite{fernando} and Einstein-aether theory \cite{jacobson} are other proposals of Lorentz violation. The SME is experimentally accessible for studying possible observable signals of violation the particle Lorentz symmetry. In this model, the local Lorentz violating is provided by a spontaneous symmetry breaking potential due to self-interacting tensor fields, whose vacuum expectation value (VEV) yields to background tensor fields. A specific theory is the bumblebee field $B^\mu$, an self-interacting tensor field has a nonzero VEV $b^\mu$ which defines a privileged direction in spacetime and spontaneously breaks the local Lorentz symmetry.

There are many forms of the bumblebee potential. One is the smooth functional bumblebee potential $V(X)=kX^2/2$ which has a minimum when the bumblebee field equals its VEV, i.e., $V=0$ and $V'(X)=kX=0$ if $X=0$, where $k$ is a constant \cite{bluhm} and $X=B^\mu B_\mu\pm b^2$. With this potential, the four dimensional static black hole solutions were found by Bertolami {\it et al.} \cite{bertolami} and by Casana {\it et al.} \cite{casana}; the higher $D$-dimensional static black hole was found by Ding {\it et al.} \cite{ding2022}; the rotating black hole solutions were built by Ding {\it et al.} \cite{ding2020} and by Jha {\it et al.} \cite{jha}. An other form of bumblebee potential is a linear multiplier function $V(X)=\lambda X/2$, where $\lambda$ is a Lagrange-multiplier. With this potential, the Schwarzschild-anti-de Sitter-like and Schwarzschild-de Sitter-like black hole solutions were found by Maluf {\it et al.} \cite{maluf}, \textcolor[rgb]{0.00,0.00,1.00}{wherein the radial and tangential pressures are different, like as an anisotropic fluid.}

The quantum field theory in a spacetime with a horizon exhibits thermal behavior which must be thought of as an emergent phenomena \cite{padmanabhan}. A black hole event horizon has the conceptions of temperature and entropy \cite{hawing1975}, and the first law of thermodynamics $dU=TdS-PdV$ can be built, where $P$ is the radial pressure of the matter, $U$ is the internal energy and $V$ is the volume of the black hole. There also has transfer of heat energy and the conception of phase transition \cite{hawking}. Then the gravity system can be mapped to a thermodynamic system and studying on black hole thermodynamics can help to construct a quantum theory of gravity.

We will study  the high dimensional black hole solutions and its thermodynamics and phase transition with a negative cosmological constant in  the theory of Einstein-bumblebee gravity. The rest paper is organized as follows.   In Sec. II we give the background for the \textcolor[rgb]{0.00,0.00,1.00}{Einstein-bumblebee} theory and derive the  black hole solution by solving the gravitational field equations. In Sec. III, we study its thermodynamical properties. In Sec. IV, we study the phase transitions and find some effects of the Lorentz breaking constant $\ell$.  Sec. V is for a summary.

\section{High dimensional AdS-like Black hole in Einstein-bumblebee gravity}

In the bumblebee gravity model, one introduces the bumblebee vector field $B_{\mu}$ which has a nonzero vacuum expectation value,  \textcolor[rgb]{0.00,0.00,1.00}{leading to} a spontaneous Lorentz symmetry breaking in the gravitational sector via a given potential. In the higher $D\geq4$ dimensional spacetime, the
 action of Einstein-bumblebee gravity is \cite{ding2021},
\begin{eqnarray}
\mathcal{S}=
\int d^Dx\sqrt{-g}\Big[\frac{R-2\Lambda}{2\kappa}+\frac{\varrho}{2\kappa} B^{\mu}B^{\nu}R_{\mu\nu}-\frac{1}{4}B^{\mu\nu}B_{\mu\nu}
-V(B_\mu B^{\mu}\mp b^2)+\mathcal{L}_M\Big], \label{action}
\end{eqnarray}
where $R$ is Ricci scalar,  $\Lambda$ is an negative cosmological constant. $\kappa=8\pi G_D/c^4$, where $G_D$ is the $D$-dimensional gravitational constant and has the relation to Newton's constant $G$ as $G_D=G\Omega_{D-2}/4\pi$ \cite{boulware}, $\Omega_{D-2}=2\sqrt{\pi}^{D-1}/\Gamma[(D-1)/2]$ is the area of a unit $(D-2)$-sphere. Here and hereafter, we take $G_D=1$ and $c=1$ for convenience.

The coupling constant $\varrho$ dominates the non-minimal gravity interaction to bumblebee field $B_\mu$. The term $\mathcal{L}_M$ represents possible interactions with matter or external currents. The potential $V(B_\mu B^{\mu}\mp b^2)$ triggers Lorentz and/or $CPT$ (charge, parity and time) violation, where the field $B_{\mu}$ acquires a nonzero VEV, $\langle B^{\mu}\rangle= b^{\mu}$, satisfying the condition $B^{\mu}B_{\mu}\pm b^2=0$. The constant $b$ is a real positive constant. Another vector $b^{\mu}$ is a function of the spacetime coordinates and has a constant value $b_{\mu}b^{\mu}=\mp b^2$, where $\pm$ signs mean that $b^{\mu}$ is timelike or spacelike, respectively. It gives a nonzero vacuum expectation value (VEV) for bumblebee field $B_{\mu}$ indicating that the vacuum of this model obtains a prior direction in the spacetime.
The bumblebee field strength is
\begin{eqnarray}
B_{\mu\nu}=\partial_{\mu}B_{\nu}-\partial_{\nu}B_{\mu}.
\end{eqnarray}
This antisymmetry of $B_{\mu\nu}$ implies the constraint \cite{bluhm}
\begin{eqnarray}
\nabla ^\mu\nabla^\nu B_{\mu\nu}=0.
\end{eqnarray}

Varying the action (\ref{action}) with respect to the metric yields the gravitational field equations
\begin{eqnarray}\label{einstein0}
G_{\mu\nu}+\Lambda g_{\mu\nu}=\kappa T_{\mu\nu}^B+\kappa T_{\mu\nu}^M,
\end{eqnarray}
where $G_{\mu\nu}=R_{\mu\nu}-g_{\mu\nu}R/2$, and the bumblebee energy momentum tensor $T_{\mu\nu}^B$ is
\begin{eqnarray}\label{momentum}
&&T_{\mu\nu}^B=B_{\mu\alpha}B^{\alpha}_{\;\nu}-\frac{1}{4}g_{\mu\nu} B^{\alpha\beta}B_{\alpha\beta}- g_{\mu\nu}V+
2B_{\mu}B_{\nu}V'\nonumber\\
&&+\frac{\varrho}{\kappa}\Big[\frac{1}{2}g_{\mu\nu}B^{\alpha}B^{\beta}R_{\alpha\beta}
-B_{\mu}B^{\alpha}R_{\alpha\nu}-B_{\nu}B^{\alpha}R_{\alpha\mu}\nonumber\\
&&+\frac{1}{2}\nabla_{\alpha}\nabla_{\mu}(B^{\alpha}B_{\nu})
+\frac{1}{2}\nabla_{\alpha}\nabla_{\nu}(B^{\alpha}B_{\mu})
-\frac{1}{2}\nabla^2(B^{\mu}B_{\nu})-\frac{1}{2}
g_{\mu\nu}\nabla_{\alpha}\nabla_{\beta}(B^{\alpha}B^{\beta})\Big].
\end{eqnarray}
The prime denotes differentiation with respect to the argument,
\begin{eqnarray}
V'=\frac{\partial V(x)}{\partial x}\Big|_{x=B^{\mu}B_{\mu}\pm b^2}.
\end{eqnarray}
Varying instead with respect to the the bumblebee field generates the bumblebee equations of motion (supposing that there is no coupling between the bumblebee field and $\mathcal{L}_M$),
\begin{eqnarray}\label{motion}
\nabla ^{\mu}B_{\mu\nu}=2V'B_\nu-\frac{\varrho}{\kappa}B^{\mu}R_{\mu\nu}.
\end{eqnarray}

The contracted Bianchi identities ($\nabla ^\mu G_{\mu\nu}=0$) lead to conservation of the total energy-momentum tensor
\begin{eqnarray}\label{}
\nabla ^\mu T_{\mu\nu}=\nabla ^\mu\big( T^B_{\mu\nu}+T^M_{\mu\nu}-\Lambda g_{\mu\nu}\big)=0.
\end{eqnarray}
We suppose that there is no matter field and the bumblebee field is frozen at its VEV like in Refs \cite{casana,bertolami}, i.e., it is
\begin{eqnarray}\label{bbu}
B_\mu=b_\mu.
\end{eqnarray}

For the forms of potential $V$, there are both classes: the smooth functionals of $x$, i.e., $\kappa x^2/2$, that minimized by the condition $x=0$ and, the Lagrange-multiplier form $\lambda x/2$, where $\lambda$ is a non-zero constant and deserves as a Lagrange-multiplier field \cite{bluhm}. The smooth potential
does not admit new black hole solution with nonzero cosmological constant $\Lambda$ \cite{maluf},
 \textcolor[rgb]{0.00,0.00,1.00}{one can also see the proof in the appendix for more details}. 

In order to generate a black hole solution in the case of non-zero cosmological constant $\Lambda$, we should apply a linear function form of  the Lagrange-multiplier potential
 \begin{eqnarray}
V=\frac{\lambda}{2}(B_\mu B^\mu-b^2).\label{potential}
\end{eqnarray}
This potential is $V=0$ under the condition (\ref{bbu}) and its derivative is $V'=\lambda/2$ which can modify the Einstein equations. However, this additional degree of freedom of the $\lambda$ field is auxiliary. Then the first two terms in Eq. (\ref{momentum}) are like those of the electromagnetic field, the only distinctiveness are the coupling items to Ricci tensor and this $\lambda$ term. Under this condition,  Eq. (\ref{einstein0}) leads to gravitational field equations \cite{ding2021}
\begin{eqnarray}\label{bar}
G_{\mu\nu}+\Lambda g_{\mu\nu}=\kappa (\lambda b_\mu b_\nu+b_{\mu\alpha}b^{\alpha}_{\;\nu}-\frac{1}{4}g_{\mu\nu} b^{\alpha\beta}b_{\alpha\beta})+\varrho\Big(\frac{1}{2}
g_{\mu\nu}b^{\alpha}b^{\beta}R_{\alpha\beta}- b_{\mu}b^{\alpha}R_{\alpha\nu}
-b_{\nu}b^{\alpha}R_{\alpha\mu}\Big)
+\bar B_{\mu\nu},
\end{eqnarray}
with
\begin{eqnarray}\label{barb}
&&\bar B_{\mu\nu}=\frac{\varrho}{2}\Big[
\nabla_{\alpha}\nabla_{\mu}(b^{\alpha}b_{\nu})
+\nabla_{\alpha}\nabla_{\nu}(b^{\alpha}b_{\mu})
-\nabla^2(b_{\mu}b_{\nu})-g_{\mu\nu}\nabla_\alpha\nabla_\beta(b^\alpha b^\beta)\Big].
\end{eqnarray}
The static spherically symmetric black hole metric in the $D$ dimensional spacetime  has the form\begin{eqnarray}\label{metric}
&&ds^2=-e^{2\phi(r)}dt^2+e^{2\psi(r)}dr^2+r^2d\Omega_{D-2}^2,
\end{eqnarray}
where $\Omega_{D-2}$ is a standard $D-2$ sphere. In this static spherically symmetric spacetime, the most general form for the bumblebee field would be $b_\mu=(b_t,b_r,0,0)$, where $b_t$ and $b_r$ are functions of $r$ subject to the constraint $-b_t^2 e^{ -2\phi}+ b_r^2 e^{-2 \psi} = b^2$, here $b$ is a positive constant. In this general case, the bumblebee field has both radial and a time component for the vacuum expectation value. \textcolor[rgb]{0.00,0.00,1.00}{In the purely radial case $b_t=0$, the authors in Refs. \cite{bertolami,casana} obtained new black hole solutions indeed. But for the general case(temporal and radial), the authors in Ref. \cite{bertolami} obtain a slightly perturbed metric, where one cannot constrain the physical parameters from the observed limits on the PPN(parameterized post-Newtonian) parameters. Hence here we consider only the purely radial case to get a black hole solution and let the general case for the future work.}

We pay attention to that the bumblebee field has a radial vacuum energy expectation because that the spacetime curvature has a strong radial variation, on the contrary that the temporal changes are very slow. Now the bumblebee field is supposed to be spacelike  as that
\begin{eqnarray}\label{bu}
b_\mu=\big(0,be^{\psi(r)},0,0,\cdots,0\big).
\end{eqnarray}
 Then the bumblebee field strength is
\begin{eqnarray}
b_{\mu\nu}=\partial_{\mu}b_{\nu}-\partial_{\nu}b_{\mu},
\end{eqnarray}
whose components are all zero. And their divergences are all zero, i.e.,
\begin{eqnarray}
\nabla^{\mu}b_{\mu\nu}=0.
\end{eqnarray}
From the bumblebee field motion equation (\ref{motion}), we have the projection of the Ricci tensor along the bumblebee field is
\begin{eqnarray}
b^{\mu}R_{\mu\nu}=\frac{\kappa\lambda}{\varrho}b_\nu\label{motion2}.
\end{eqnarray}

As to gravitational field equation (\ref{bar}), one can obtain the following three  component equations
\begin{eqnarray}
&&(D-2)(1+\ell)\big[2r\psi'-(D-3)\big]+e^{2\psi}\big[(D-2)(D-3)-2\Lambda r^2\big]=0,\label{tt}\\
&&2\ell r^2(\phi''+\phi'^2-\phi'\psi')-2\ell(D-2)r(\psi'+\phi')-2(D-2)r\phi'\nonumber\\
&&\qquad\qquad +e^{2\psi}\big[(D-2)(D-3)+2\kappa \lambda b^2r^2-2\Lambda r^2\big]-(1+\ell)(D-2)(D-3)=0,\label{rr}\\
&&(1+\ell)\Big[r^2(\phi''+\phi'^2-\phi'\psi')+\frac{(D-3)(D-4)}{2}+(D-3)r(\phi'-\psi')\Big]\nonumber\\
&&\qquad\qquad+
e^{2\psi}\Big[\Lambda r^2-\frac{(D-3)(D-4)}{2}\Big]=0\label{theta},
\end{eqnarray}
where we have redefined the Lorentz-violating parameter $\ell=\varrho b^2$ and, the prime $'$ is the derivative with respect to the corresponding argument, respectively.
From the Eq. (\ref{tt}), one can obtain a metric function as
\begin{eqnarray}\label{}
&&e^{2\psi}=\frac{1+\ell}{f(r)},
\end{eqnarray}
where
\begin{eqnarray}\label{metricfun}
&&f(r)=1-\frac{16\pi M}{(D-2)\Omega_{D-2}r^{D-3}}-\frac{2\Lambda}{(D-1)(D-2)}r^2,
\end{eqnarray}
and  $M$ is the mass of the black hole. In order to recover the 4-dimensional Schwarzschild-like solution \cite{casana} in the cases of $\Lambda=0$, then we let another function to\begin{eqnarray}\label{}
&&e^{2\phi}=f(r).
\end{eqnarray}
It is easy to prove that the above results of $e^{2\phi}=f(r),e^{2\psi}=(1+\ell)/f(r)$ can meet the gravitational Eq. (\ref{theta}). Then the bumblebee field is \begin{eqnarray}\label{bu}
b_\mu=\big(0,b\sqrt{(1+\ell)/f(r)},0,0,\cdots,0\big).
\end{eqnarray}

With the bumblebee field motion equation (\ref{motion2}) or the gravitational Eq. (\ref{rr}),  a solution of this type will be possible if and only if
\begin{eqnarray}\label{constraint}
&&\Lambda=\frac{(D-2)\kappa\lambda}{2\varrho}(1+\ell),
\end{eqnarray}
 which is the constraint on the additional field $\lambda$ coming from the potential (\ref{potential}). It is easy to see that if the Lagrange-multiplier field $\lambda=0$, then $\Lambda=0$ and the metric (\ref{metricfun}) is the $D$-dimensional Schwarzschild-like black hole \cite{ding2022}. So $\lambda$ isn't a new freedom and we can define the effective cosmological constant $\Lambda_e$ as following to absorb it,
\begin{eqnarray}\label{}
&&\Lambda_e=\frac{(D-2)\kappa\lambda}{2\varrho}.
\end{eqnarray}
Then $\Lambda=(1+\ell)\Lambda_e$ and,
 the present black hole metric is
\begin{eqnarray}\label{metricl}
&&ds^2=-f(r)dt^2+\frac{1+\ell}{f(r)}dr^2+r^2d\Omega_{D-2}^2,
\end{eqnarray} with the metric function $f(r)$ as
\begin{eqnarray}\label{}
&&f(r)=1-\frac{16\pi M}{(D-2)\Omega_{D-2}r^{D-3}}-\frac{2(1+\ell)\Lambda_e}{(D-1)(D-2)}r^2.
\end{eqnarray}
When LV constant $\ell\rightarrow0$, the effective cosmological constant $\Lambda_e\rightarrow \Lambda$ and it recovers the  Schwarzschild-AdS black hole.
\textcolor[rgb]{0.00,0.00,1.00}{When $r\rightarrow\infty$, the metric function
\begin{eqnarray}\label{}
&&f(r)=1-\frac{2(1+\ell)\Lambda_e}{(D-1)(D-2)}r^2,
\end{eqnarray}
and the metric (\ref{metricl}) becomes a $D$-dimensional AdS-like spacetime.}
The black hole horizon locates at the largest real root of the following equation
\begin{eqnarray}\label{horizon}
&&1-\frac{16\pi M}{(D-2)\Omega_{D-2}r_h^{D-3}}-\frac{2(1+\ell)\Lambda_e}{(D-1)(D-2)}r_h^2=0.
\end{eqnarray}
It is easy to see that it can give the black hole mass $M$ from the horizon radius,
\begin{eqnarray}\label{hmass}
&&M=\frac{(D-2)\Omega_{D-2}}{16\pi }r_h^{D-3}-\frac{(1+\ell)\Lambda_e}{8\pi(D-1)}\Omega_{D-2}r_h^{(D-1)}.
\end{eqnarray}
It shows that the bumblebee field affects the location of the event horizons, contrary to the Schwarzschild-like solution in which the horizon radius is the same as of the Schwarzschild black hole.

\section{Thermodynamics of Einstein-bumblebee AdS-like black hole}
A black hole is not only a gravity system, but also a special thermodynamic system due to that its surface gravity $\kappa$ and horizon area $A$ are the close similarity to the temperature $T$ and entropy $S$, $T=\kappa/2\pi$, $S=A/4$. Four laws of black hole thermodynamics were established in Ref. \cite{bardeen}. LV can modify the geometry of a black hole, so in this kind of black hole, can the thermodynamical laws still hold? In this section we study the Smarr formula and the first law of this LV black hole.
\subsection{Smarr formula}
Suppose that $\mathcal{M}$ is this $D$-dimensional spacetime satisfying the Einstein equations, \textcolor[rgb]{0.00,0.00,1.00}{$\xi^a=(1,0,0,0,\cdots,0)$ is a Killing vector on $\mathcal{M}$, timelike near infinity. In $\mathcal{M}$,  there is a spacelike hypersurface $S$ with a co-dimension 2-surface boundary $\partial S$, and $\xi^a$ is normal to the $S$}. The boundary $\partial S$ has two components: an inner boundary at the event horizon $\partial S_h$ and an outer boundary at infinity $\partial S_\infty$. In order to construct a Komar integral relation for non-zero cosmological constant, one should introduce Killing potential $\omega^{ab}$ \cite{kastor} which can be obtained according to relation $\xi^b=\nabla_a\omega^{ab}$. For the static Killing vector $\xi^a$, we have
\begin{eqnarray}\label{}
\omega^{rt}=-\omega^{tr}=\frac{r}{D-1}.
\end{eqnarray}
We can integrate the Killing equation $\nabla_b(\nabla^b\xi^a)=-R^a_b\xi^b$ over a hypersurface $S$,
\begin{eqnarray}\label{intone}
\int_{\partial S}\nabla^b\xi^ad\Sigma_{ab}=-\int_SR^a_b\xi^bd\Sigma_a,
\end{eqnarray}
where $d\Sigma_{ab}$ and $d\Sigma_{a}$ are the surface elements of $\partial S$ and $S$, respectively.
With the metric (\ref{metric}) in mind, we have the relation
\begin{eqnarray}\label{}
R^a_b\xi^b=\frac{2}{D-2}(\Lambda_e,0,0,0,\cdots,0)=\frac{2\Lambda_e}{D-2}\xi^a,
\end{eqnarray}
then the Eq. (\ref{intone}) can be rewritten as
\begin{eqnarray}\label{inttwo}
\frac{D-2}{8\pi}\int_{\partial S}\left(\nabla^b\xi^a+\frac{2}{D-2}\Lambda_e\omega^{ab}\right)d\Sigma_{ab}=0,
\end{eqnarray}
which is multiplied by the normalization factor $(D-2)/8\pi$ and called the Komar integral relation.
The non-vanishing components of the tensor $\nabla^b\xi^a$ are given by
\begin{eqnarray}\label{}
\nabla^r\xi^t=-\nabla^t\xi^r=\frac{1}{(1+\ell)(D-2)}\left[\frac{(D-3)8\pi M}{\Omega_{D-2}r^{D-2}}-\frac{2(1+\ell)\Lambda_er}{(D-1)}\right].
\end{eqnarray}
The closed 2-surface $\partial S$ has two parts, horizon $\partial S_h$ and infinite $\partial S_\infty$, so Eq. (\ref{inttwo}) can be rewritten as
\begin{eqnarray}\label{intthree}
\frac{D-2}{8\pi}\int_{\partial S_\infty}\left(\nabla^b\xi^a+\frac{2}{D-2}\Lambda_e\omega^{ab}\right)d\Sigma_{ab}=\frac{D-2}{8\pi}\int_{\partial S_h}\left(\nabla^b\xi^a+\frac{2}{D-2}\Lambda_e\omega^{ab}\right)d\Sigma_{ab}.
\end{eqnarray}
If we use the 2-surface element $d\Sigma_{rt}=-\sqrt{1+\ell}r^{D-2}d\Omega_{D-2}/2$, which is slightly modified by the factor $\sqrt{1+\ell}$, the left hand side of this integral is
\begin{eqnarray}\label{}
\frac{D-2}{8\pi}\int_{\partial S_\infty}\left(\nabla^b\xi^a+\frac{2}{D-2}\Lambda_e\omega^{ab}\right)d\Sigma_{ab}
=-\frac{(D-3)M}{\sqrt{1+\ell}};
\end{eqnarray}
the second term of its right hand side is
\begin{eqnarray}\label{}
\frac{\Lambda_e}{8\pi}\int_{\partial S_h}2\omega^{ab}d\Sigma_{ab}
=\frac{2PV}{\sqrt{1+\ell}},
\end{eqnarray}
where the pressure $P$ and the thermodynamic volume $V$ are
\begin{eqnarray}\label{vol}
P=-\frac{\Lambda_e}{8\pi},\;\;V=(1+\ell)\frac{\Omega_{D-2}}{D-1}r_h^{D-1};
\end{eqnarray}
and the first term of its right hand side is
\begin{eqnarray}\label{}
\frac{D-2}{8\pi}\int_{\partial S_h}\nabla^b\xi^ad\Sigma_{ab}
=-\frac{(D-2)TS}{\sqrt{1+\ell}},
\end{eqnarray}
where the temperature $T$ and entropy $S$ are
\begin{eqnarray}\label{area}
T=\frac{4}{(D-2)\sqrt{1+\ell}r_h}\left[\frac{(D-3) M}{\Omega_{D-2}r_h^{D-3}}-\frac{(1+\ell)\Lambda_er_h^2}{4\pi(D-1)}\right],
\;\;S=\frac{A}{4}=\frac{\sqrt{1+\ell}}{4}\Omega_{D-2}r_h^{D-2}.
\end{eqnarray}
Lastly, the integral (\ref{intthree}) can give the Smarr formula
\begin{eqnarray}\label{}
(D-3)M=(D-2)TS-2PV.
\end{eqnarray}
One can see that the Smarr formula can still be constructed in this LV black hole spacetime as long as its temperature, entropy and volume are slightly modified.

Why the thermodynamic volume $V$ (\ref{vol}) and area $A$ (\ref{area}) of the 2-surface $\partial S $ is different from those in Ref. \cite{kubiznak2017} \textcolor[rgb]{0.00,0.00,1.00}{(see the Eqs. 2.25 and 2.28 in it for 4-dimensions)} by the factor $\sqrt{1+\ell}$ and $(1+\ell)$?
Because these two quantities characterize a spacetime and should be entirely derived from thermodynamic considerations.  So one should firstly determine its temperature. The temperature $T$ should be given by the black hole's thermodynamic process--- its Hawking radiation process \cite{ding2008}
\begin{eqnarray}\label{tem}
T=\frac{\sqrt{f'(r_h)g'(r_h)}}{4\pi}
=\frac{1}{4\pi\sqrt{1+\ell}\;r_h}\left[(D-3)-\frac{2}{D-2}(1+\ell)\Lambda_er_h^2\right],
\end{eqnarray}
\textcolor[rgb]{0.00,0.00,1.00}{where $g(r)=f(r)/(1+\ell)$.}

For the thermodynamic volume $V$, it is the quantity thermodynamically conjugate to the pressure $P$, then it should be \textcolor[rgb]{0.00,0.00,1.00}{(see the formula 3.1 in Ref.  \cite{kubiznak2017})}
 \begin{eqnarray}\label{volume}
V\equiv\Big(\frac{\partial M}{\partial P}\Big)_S=(1+\ell)\frac{\Omega_{D-2}}{D-1}r_h^{D-1},
\end{eqnarray}
with the  Eq. (\ref{hmass}) for the mass $M$.
For the thermodynamic area $A$, it is the quantity thermodynamically conjugate to the temperature $T$, then it should be
 \begin{eqnarray}\label{entropy}
A=4S\equiv4\int\frac{dM}{T}=\sqrt{1+\ell}\Omega_{D-2} r_h^{D-2}.
\end{eqnarray}
\textcolor[rgb]{0.00,0.00,1.00}{Therefore, one can see that the LV has affected the known spacetime geometry by redefinition of the geometrical area and volume.}

\subsection{The first law}
\textcolor[rgb]{0.00,0.00,1.00}{Nextly, we derive the first law of black hole thermodynamics via Hamiltonian perturbation method}. Note that in Ref. \cite{kastor}, the authors  used this method to lead to the first law and they found it was equivalent to \textcolor[rgb]{0.00,0.00,1.00}{the derivative of $M$  method which will be appeared in the appendix B}, where $M$ is determined by Eq. (\ref{hmass}).

In $D$-dimensional manifold $\mathcal{M}$ with metric $g_{ab}$ of signature $(-,+,\cdots,+)$, let $\Sigma$ be a family of $(D-1)$-dimensional spacelike submanifolds with unite timelike normal field $n_a$ and induced metric $s_{ab}$, i.e.,
\begin{eqnarray}\label{}
g_{ab}=-n_an_b+s_{ab},\qquad n_cn^c=-1,\qquad n_cs^{cb}=0.
\end{eqnarray}
The Hamiltonian variables are $s_{ab}$ and its conjugate momentum $\pi_{ab}$. The energy density $\rho=T_{ab}n^an^b$  and momentum density $J_a=T_{bc}n^bs^c_{\;a}$ must satisfy the Hamiltonian and momentum constrain equations,
\begin{eqnarray}\label{constrain}
H=16\pi G\rho,\qquad H_a=-16\pi GJ_a,
\end{eqnarray}
where
\begin{eqnarray}\label{ha}
H=-2G_{ab}n^an^b=-\left[\frac{(D-2)(D-3)\ell}{(1+\ell)r^2}+2\Lambda_e\right],\qquad H_a=-2G_{bc}n^bs^c_{\;a}=0
\end{eqnarray}
for the given case. Let the Killing vector $\xi ^a=Fn^a+\beta^a$ with $n_c\beta^c=0$. The Hamiltonian density $\mathcal{H}$ for evolution along $\xi ^a$ is given by
\begin{eqnarray}\label{}
\mathcal{H}=\sqrt{s}\left\{F\left[H+2\Lambda_e+\frac{(D-2)(D-3)\ell}{(1+\ell)r^2}\right]+\beta_aH^a\right\}.
\end{eqnarray}
We assume that the linear approximation metric $\tilde{g}_{ab} = g_{ab} + \delta g_{ab}$ is  another nearby solution to the Einstein equations with a perturbed cosmological constant $\Lambda +\delta\Lambda$. The induced spatial metric and momentum for this perturbed metric are $\tilde{s}_{ab}=s_{ab}+h_{ab}$ and $\tilde{\pi}_{ab}=\pi_{ab}+p_{ab}$, where $h_{ab}=\delta s_{ab}, p_{ab}=\delta \pi_{ab}$. From Hamilton's equations for the zeroth order, the linearized constrain operators $\delta H$ and $\delta H_a$ combine to form a total derivative,
\begin{eqnarray}\label{}
F\delta H +\beta^a \delta H_a =-D_cB^c,
\end{eqnarray}
where $D_a$ is the covariant derivative operator on $\Sigma$ compatible with $s_{ab}$, and \footnote{ In the general case of non-vanishing extrinsic curvature, the boundary vector $B^a$
includes the term \cite{kastor2014} $\frac{1}{\sqrt{s}}\beta ^b(\pi^{cd}h_{cd}s^a_{\;b}-2\pi^{ac}h_{bc}-2p^a_{\;b})$.}
\begin{eqnarray}\label{}
B^a=F(D^ah-D_bh^{ab})-hD^aF+h^{ab}D_bF.
\end{eqnarray}
The linear constraints (\ref{constrain}) take the form of a Gauss law \footnote{From the Eq. (\ref{ha}), the perturbation $\delta H=16\pi G\delta\rho\sim -2\delta\Lambda_e-\delta(1/r^2)\sim-2\delta\Lambda_e $ at large distance. So the contribution of this additional matter field source, i.e., the purely radial bumblebee field is only included in the term $\delta\Lambda_e$.}
\begin{eqnarray}\label{gauss}
D_cB^c=2F\delta \Lambda_e.
\end{eqnarray}
Using Killing potential $\omega^{ab}$, we have $F=-n_a\xi^a=-D_c(n_a\omega^{ca})$, and rewrite Eq. (\ref{gauss}) in the integral form,
\begin{eqnarray}\label{integral}
I=\int_{\partial\Sigma}da_c(B^c+2\omega^{cd}n_d\delta \Lambda_e)=0,
\end{eqnarray}
which is taken to extend from a boundary $\partial \Sigma_h$ at the bifurcation sphere of the horizon to a boundary $\partial\Sigma_\infty$ at infinity, i.e., $I_\infty-I_h=0$,
\begin{eqnarray}\label{}
\int_{\partial\Sigma_\infty} da_c(B^c+2\omega^{cd}n_d\delta \Lambda_e)-\int_{\partial\Sigma_h} da_c(B^c+2\omega^{cd}n_d\delta \Lambda_e)=0.
\end{eqnarray}
We choose $F\simeq\sqrt{f},\;n_a=-F\nabla_at,\;da_r=\sqrt{1+\ell}\;r^{D-2}d\Omega_{D-2}/\sqrt{f}$, and
\begin{eqnarray}\label{}
h_{rr}\simeq\delta\left(\frac{1+\ell}{f}\right),\qquad \delta f=-\left[\frac{16\pi\delta M}{(D-2)\Omega_{D-2}r^{D-3}}+\frac{2(1+\ell)\delta \Lambda_er^2}{(D-1)(D-2)}\right],
\end{eqnarray}
then the integral (\ref{integral}) can be simplified by
\begin{eqnarray}\label{}
I=\frac{1}{\sqrt{1+\ell}}\int_{\partial\Sigma} r^{D-2}d\Omega_{D-2}\left[\frac{D-2}{r}\delta f-\frac{f'\delta A}{\sqrt{1+\ell}A}+\frac{2(1+\ell)r\delta \Lambda_e}{(D-1)}\right],
\end{eqnarray}
where the prime $'$ denotes the derive with the argument, and $A$ is the area of the boundary. So $I_\infty=-16\pi\delta M/\sqrt{1+\ell}$ since $1/A\rightarrow0$ at infinite, and
\begin{eqnarray}\label{}
I_h=\frac{16\pi}{\sqrt{1+\ell}}[-T\delta S-V\delta P],
\end{eqnarray}
since $f(r_h)=0,\;f'(r_h)=4\pi\sqrt{1+\ell}T,\;\Lambda_e=-8\pi P,\;A=4S$ and $V=(1+\ell)\Omega_{D-2}r_h^{D-1}/(D-1)$ \footnote{For the general form of the volume $V$, see the formula (22) in Ref. \cite{kastor}}. Lastly, the first law is obtained,
\begin{eqnarray}\label{}
\delta M=T\delta S+V\delta P,
\end{eqnarray}
which still also holds in this LV black hole spacetime.

\section{Phase transition of Einstein-bumblebee AdS-like black hole}
A phase transition is a discontinuous change in the properties of a substance, as its environment is changed only infinitesimally. An isolated Schwarzschild black hole is thermodynamically unstable due to its negative heat capacity. However for a Schwarzschild AdS black hole, besides the negative heat phase, a phase of positive heat capacity will occur. At a given temperature and pressure, the stable phase is always the one with the lower Gibbs free energy. In an AdS space, the Gibbs energy of the thermal radiation phase is zero \cite{wei}.  So the negative free energy of the positive heat capacity black hole phase is stable phase. In this section, we study this two phase transitions.

From the Eq. (\ref{horizon}), we can express the mass $M$ as
\begin{eqnarray}\label{}
M=\frac{(D-2)\Omega_{D-2}}{16\pi}r_h^{D-3}+PV.
\end{eqnarray}
$M$ can be interpreted as an enthalpy \cite{kastor} $H=U+PV$, so the black hole internal energy $U$ is
\begin{eqnarray}\label{}
U=\frac{(D-2)\Omega_{D-2}}{16\pi}r_h^{D-3}.
\end{eqnarray}
Then the first law (\ref{first}) can be rewritten as following
\begin{eqnarray}\label{}
dU=TdS-PdV,
\end{eqnarray}
which is the same as the fundamental thermodynamical equation of a simple gas system.
The temperature (\ref{tem}) can be rewritten as
\begin{eqnarray}\label{}
T=\frac{1}{4\pi\sqrt{1+\ell}\;r_h}
\left[(D-3)+\frac{16\pi P}{D-2}(1+\ell)r_h^2\right],
\end{eqnarray}
which is shown in Fig. 1 when $D=4$. \begin{figure}[ht]\label{g00}
\begin{center}
\includegraphics[width=8.0cm]{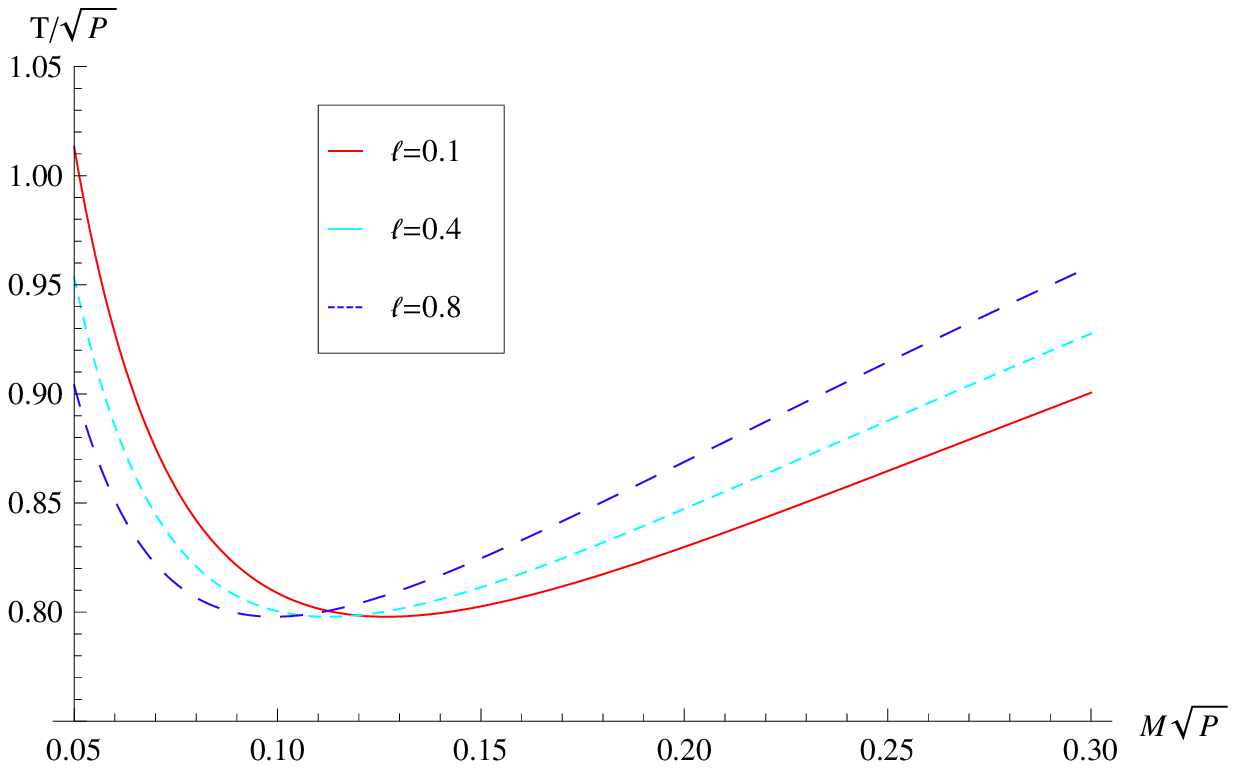}\;\;\;\includegraphics[width=8.0cm]{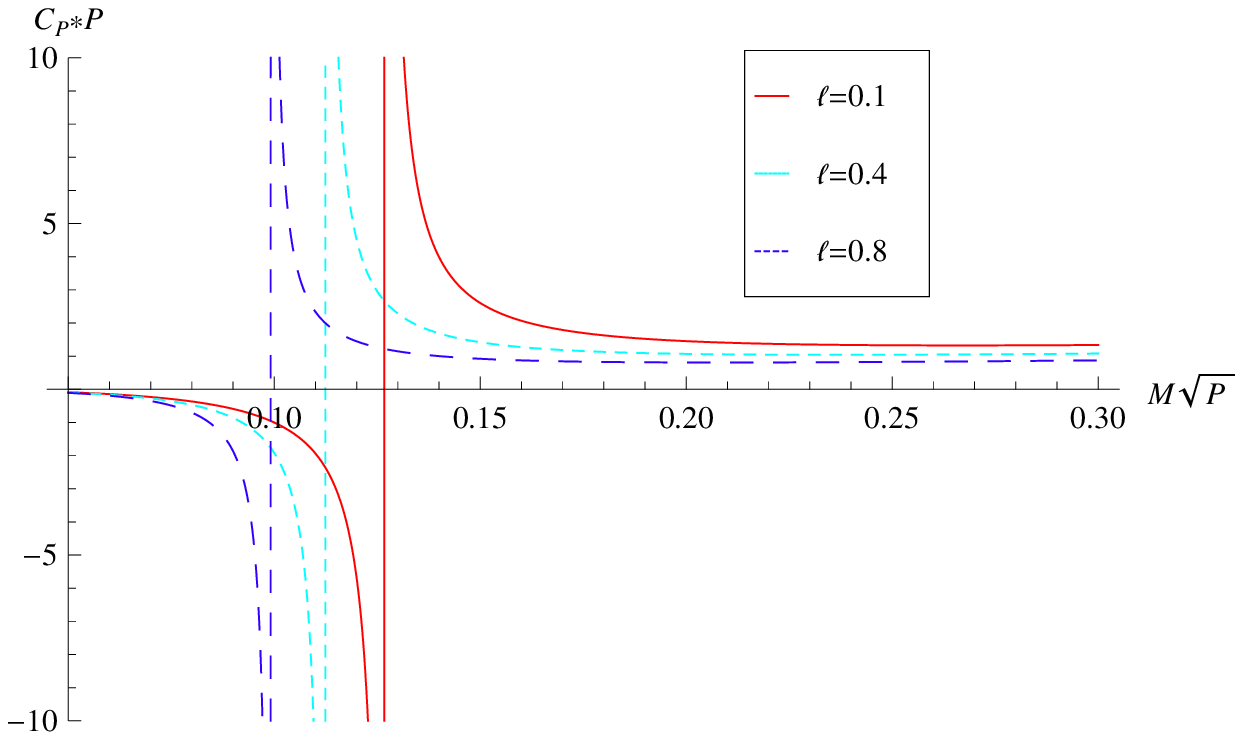}
\caption{Black hole temperature and heat capacity $C_P$ at constant pressure $P$ vs. black hole mass $M$ with $D=4$ and different coupling constant $\ell$. The temperature has a minimum value and the the heat capacity is divergent at the point of black hole mass $M=M_d$, where $M_{d}=1/\sqrt{18(1+\ell)\pi P}$.}
\end{center}
\end{figure}
There exist a minimum temperature
\begin{eqnarray}\label{}
T_{min}=2\sqrt{\frac{(D-3)P}{(D-2)\pi}},
\end{eqnarray}
when $r_h=\sqrt{(D-2)(D-3)/[16(1+\ell)\pi P]}$. Note that $T_{min}$ isn't dependent on $\ell$. From the left panel of the Fig. 1, when $T>T_{min}$, there are two possible black hole masses \cite{hawking}  which can be in equilibrium with thermal radiation, i.e.,  $M>M_d$ or $M<M_d$, where $M_d$ is the black hole mass whose heat capacity is divergent (\ref{md}). Fig. 1 shows that the temperature of the large black hole ($M>M_d$) increases when the LV coupling constant $\ell$ becomes bigger.
At a fixed pressure $P$, the black hole heat capacity $C_P$ is
\begin{eqnarray}\label{}
C_P=\Big(\frac{\partial H}{\partial T}\Big)_P=\frac{(D-2)\Omega_{D-2}}{4} r_h^{D-2}\frac{16\pi(1+\ell)Pr_h^2+(D-2)(D-3)}{16\pi(1+\ell)Pr^2_h-(D-2)(D-3)},
\end{eqnarray}
which is also shown in Fig. 1 when $D=4$. From the right panel of the Fig. 1, one can see that the small black hole mass has negative specific heat, so it is unstable and will decay. While the large black hole mass has positive heat capacity  and therefore it is thermally stable. It is easy to see that the heat capacity at constant pressure is divergent at the black hole mass $M_d$,
\begin{eqnarray}\label{md}
M_d=\frac{(D-2)^2\Omega_{D-2}}{8\pi(D-1)} \left[\frac{(D-2)(D-3)}{16(1+\ell)\pi P}\right]^{(D-3)/2}.
\end{eqnarray}
It is easy to see that the LV constant $\ell$ decreases the mass $M_d$, showing that the phase transition from a small black hole to large one will more easily occur.
At this divergent point, the black hole entropy is
\begin{eqnarray}\label{}
S_d=4^{(1-D)}\Omega_{D-2}\sqrt{1+\ell}^{(3-D)} \left[\frac{(D-2)(D-3)}{\pi P}\right]^{(D-2)/2}.
\end{eqnarray}
Fig. 2 shows the entropy with the temperature at unit pressure  when $D=4$.
\begin{figure}[ht]\label{g01}
\begin{center}
\includegraphics[width=10.0cm]{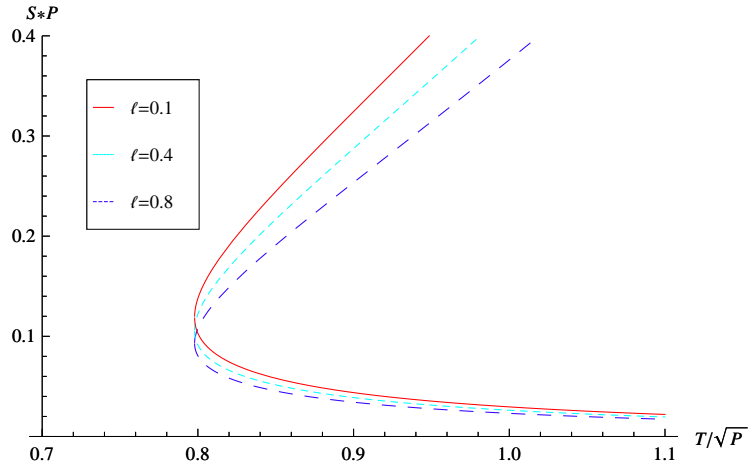}
\caption{Black hole entropy $S$ vs. black hole temperature $T$ with $D=4$ and different coupling constant $\ell$. They all have both branches, the above one is corresponding  to the large black hole $M>M_d$, the below one to the small black hole $M<M_d$. At the temperature $T=T_{min}$, the entropy is $S_d=1/8\sqrt{1+\ell}P$. At the HP temperature $T_{HP}=\sqrt{8P/3\pi}, S_{HP}=3/8\sqrt{1+\ell}P$ for the large black hole, $S_{small}=1/24\sqrt{1+\ell}P$ for the unstable small black hole. }
\end{center}
\end{figure}
 The phase transition between these two phases is called small-large black hole phase transition.

There is another kind of phase transition called Hawking-Page(HP) phase transition, which is between a metastable large black hole phase and stable large black hole phase in an anti-de Sitter spacetime. Witten \cite{witten} explained that this HP phase transition is a confinement/deconfinement phase transition in the AdS/CFT correspondence. It can also be understood as a solid/liquid phase transition in black hole chemistry \cite{kubiznak2015}. HP phase transition occurs when Gibbs free energy is zero.
The black hole Gibbs free energy can be obtained
\begin{eqnarray}\label{}
G=H-TS=\frac{\Omega_{D-2}}{8\pi (D-1)}\left[(D-2)-2\pi\sqrt{1+\ell}Tr_h\right]r_h^{D-3},
\end{eqnarray}
which is shown in Fig. 3 when $D=4$. It can be seen that there are two branches, the above parts corresponding to small black hole phase whose free energy are positive for any temperature; the below ones corresponding to metastable large black hole phase when $G>0$ and stable large black hole phase when $G<0$. When the temperature $T<T_{min}$, the spacetime has only one phase---thermal radiation,  whose free energy is zero.
\begin{figure}[ht]\label{g02}
\begin{center}
\includegraphics[width=10.0cm]{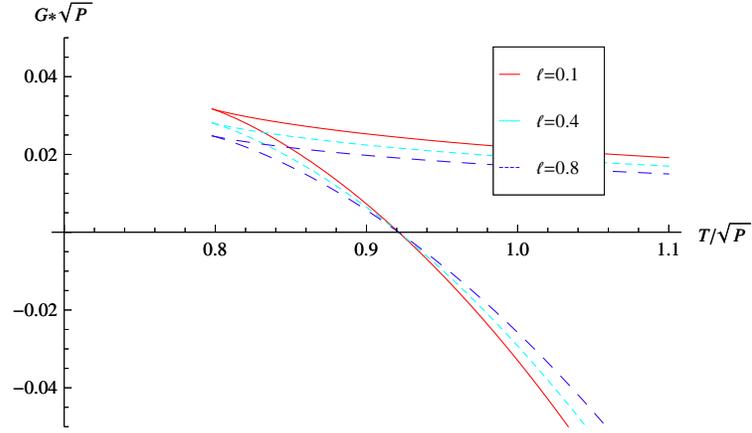}
\caption{Black hole Gibbs free energy vs. temperature with $D=4$ and different coupling constant $\ell$. They vanish at the point $T_{HP}=\sqrt{8P/3\pi}$. They all have both branches, the above one is corresponding  to the small black hole $M<M_d$, the below one to the large black hole $M>M_d$. At the temperature $T=T_{min}$, the free energy has a maximum $G_{max}=1/12\sqrt{2(1+\ell)\pi P}$.}
\end{center}
\end{figure}
At the temperature $T=T_{min}$, its free energy has a maximum value $G_{max}$,
\begin{eqnarray}\label{}
G_{max}=\frac{\Omega_{D-2}}{(D-1)\pi}2^{(3-2D)} \left[\frac{(D-2)(D-3)}{(1+\ell)\pi P}\right]^{(D-3)/2}.
\end{eqnarray}
The temperature at Hawking-Page phase transition point  can be obtained when $G=0$,
\begin{eqnarray}\label{}
T_{HP}=2\sqrt{\frac{(D-2)P}{(D-1)\pi}},
\end{eqnarray}
which is also independent on $\ell$ like the minimum temperature $T_{min}$.
At this point $r_h=\sqrt{(D-1)(D-2)/[16(1+\ell)\pi P]}$ and the black hole mass is,
\begin{eqnarray}\label{}
M_{HP}=\frac{(D-2)\Omega_{D-2}}{8\pi} \left[\frac{(D-1)(D-2)}{16(1+\ell)\pi P}\right]^{(D-3)/2},
\end{eqnarray}
its entropy is
\begin{eqnarray}\label{}
S_{HP}=4^{(1-D)}\Omega_{D-2}\sqrt{1+\ell}^{(3-D)} \left[\frac{(D-1)(D-2)}{\pi P}\right]^{(D-2)/2}.
\end{eqnarray}
We summarize the three phases: unstable small black hole, metastable large black hole and stable large black hole in Tab. \ref{tab1}; stable large black hole, metastable large black hole and thermal radiation in Fig. 4.
 \begin{table}
 \caption{Quantities of heat capacity, entropy, temperature and Gibbs free energy for three phases of the AdS black hole in Einstein-bumblebee gravity.}\label{tab1}
\begin{center}
\begin{tabular}{|c|c|c|c|c|c|}
\hline  & Black hole mass & Heat capacity & Entropy & Temperature & Free energy \\
\hline Unstable small black hole &
 $0<M<M_d$ & $C_p<0$ & $S<S_d$ & $T>T_{min}$ & $G>0$ \\
 \hline Metastable large black hole &
 $M_d<M<M_{HP}$ & $C_p>0$ & $ S_d<S<S_{HP}$ & $T_{min}<T<T_{HP}$ & $0<G<G_d$ \\
\hline Stable large black hole &
 $M>M_{HP}$ & $C_p>0$ & $S>S_{HP}$ & $T>T_{HP}$ & $G<0$ \\
\hline
\end{tabular}
\end{center}
\end{table}
\begin{figure}[ht]\label{}
\begin{center}
\includegraphics[width=10.0cm]{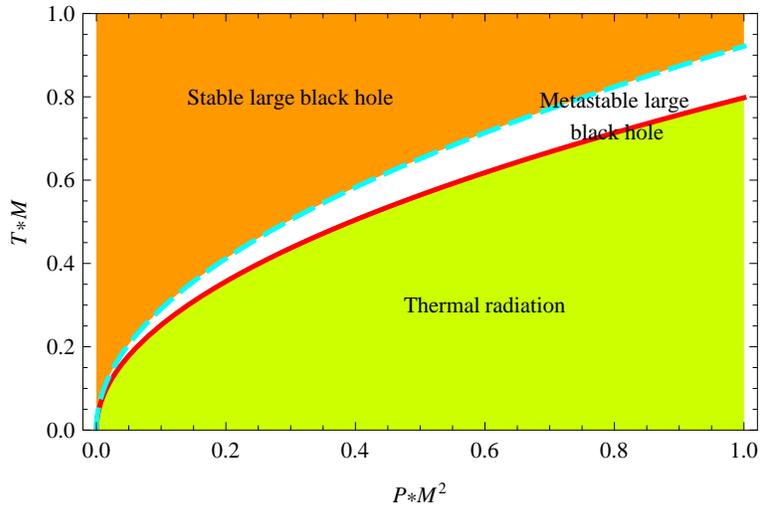}
\caption{Phase diagram ($T-P$ figure at unit mass $M$) for $D=4$ AdS black hole in Einstein-bumblebee gravity. The red solid and green dashed curves respectively correspond to the black hole minimum temperature and the HP phase transition temperature. Both curves extend to infinity.}
\end{center}
\end{figure}

From the figures 1 to 4 and the table I, one can see the following picture for the formation of a AdS black hole: i), in an AdS space, the thermal radiation with zero Gibbs free energy has a very high temperature, accidentally, a tiny mass black hole begins to form due to quantum fluctuation; ii), the tiny black hole begins to grow, its temperature decreases quickly, its free energy increases slowly; iii), when the black hole mass grows to $M_d$, its free energy approaches to a maximum, its temperature approaches to a minimum, and the small-large black hole phase transition occurs; iv), the black hole mass continues to grow, its positive free energy decreases, its temperature increases slowly and it is a metastable black hole; v), when the mass grows to $M_{HP}$, the HP phase transition occurs; vi), finally, the stable large black hole has formed.

Figure 3 also shows that the LV constant $\ell$ decreases the Gibbs free energy of the meta-stable large black hole with the mass $M_d<M<M_{HP}$, showing that the HP phase transition from a meta-stable large black hole to large stable one will be more easily to occur.

Now we consider some relationships between the both kinds of phase transition. Wei and Liu \cite{wei} found that a novel exact dual relation between the minimum temperature of the $D+1$-dimensional black hole and the HP phase transition temperature in $D$ dimensions, reminiscent of the holographic principle. Here we will examine wether it still hold or not to this LV black hole. The first relationship is for the temperature and entropy,
\begin{eqnarray}\label{}
\frac{T_{HP}}{T_{min}}=\frac{D-2}{\sqrt{(D-1)(D-3)}},\;
\frac{S_{HP}}{S_{d}}=\left(\frac{D-1}{D-3}\right)^{(D-2)/2},
\end{eqnarray}
which are only dependant on the dimension number $D$ and the same as those in Refs. \cite{wei,belhaj}.
The second relationship is between the $D$ dimensional Hawking-Page temperature $T_{HP}$ and the $D+1$ dimensional minimum temperature $T_{min}$,
\begin{eqnarray}\label{}
T_{HP}(D)=T_{min}(D+1),
\end{eqnarray}
which is  the novel dual relation found in Ref. \cite{wei}. This relation is independent of the pressure and the bumblebee coupling constant, and is like the AdS/CFT correspondence. $T_{min}$ is a physical quantity in the bulk, then $T_{HP}$ can be treated as the dual physical quantity on the boundary.
\section{Summary}

In this paper, we have derived the high dimensional AdS-like black hole solution in Einstein-bumblebee gravity theory. It can exists only under the condition that the bumblebee potential has a linear functional form with a Lagrange-multiplier field $\lambda$. This additional field is strictly constrained by the bumblebee motion equation and can be absorbed by the definition of an effective cosmological constant $\Lambda_e$. Like Schwarzschild-like black hole \cite{casana}, it can't be asymptotically to anti-de Sitter spacetime, so it is a  Schwarzschild-AdS-like black hole solution \cite{maluf}. Contrary to Schwarzschild-like black hole, the bumblebee field will affect the locations of the black hole horizon.

By using black hole chemistry, we have studied the thermodynamics and phase transitions of this high dimensional Schwarzschild-AdS-like black hole. Via Komar integral method, we find that the Smarr formula can exist as long as  its temperature, entropy and volume are slightly modified, and \textcolor[rgb]{0.00,0.00,1.00}{the LV has affected the known spacetime geometry, such as area and volume}. Via Hamiltonian perturbation method, there still exist the first law of black hole thermodynamics.

Its temperature has a minimum value $T_{min}$, corresponding to the black hole mass $M_d$. Small black hole with mass $0<M<M_d$ has a negative heat capacity $C_p$, while the large black hole with mass $M>M_d$ has a positive one. At the point $M_d$, the $C_p$ is divergent. The LV constant $\ell$ decreases the mass $M_d$, showing that the phase transition from a small black hole to large one will more easily occur.

Its Gibbs energy has a maximum value $G_{max}$ at the point of mass $M_d$ and has a zero point $G=0$ at the point of mass  $M_{HP}$. Small black hole with mass $0<M<M_d$ and large black hole with mass $M_d<M<M_{HP}$  have a positive Gibbs free energy, while the large black hole with mass $M>M_d$ has a negative one. The LV constant $\ell$ decreases the Gibbs free energy of the meta-stable large black hole with the mass $M_d<M<M_{HP}$, showing that the HP phase transition from a meta-stable large black hole to large stable one will be more easily to occur.

The dualities found in Ref. \cite{wei} still holds when the Lorentz symmetry is spontaneously broken: the ratio of temperature between small/large black hole phase transition and HP phase transition $T_{HP}/T_{min}$, and the equivalence of the temperature of $D$-dimensional HP phase transition  to the temperature of $D+1$ dimensional small/large black hole phase transition $T_{HP}(D)=T_{min}(D+1)$.

\begin{acknowledgments}  This work was supported by the Scientific Research Fund of the Hunan Provincial Education Department under No. 19A257 and No. 19A260, the National Natural Science Foundation (NNSFC)
of China (grant No. 11247013), Hunan Provincial Natural Science Foundation of China grant No. 2015JJ2085.
\end{acknowledgments}

\appendix\section{Proof for the potential $V=\kappa x^2/2$ admitting no black hole solution with $\Lambda\neq0$}
\textcolor[rgb]{0.00,0.00,1.00}{If one uses the smooth potential $V=\kappa x^2/2$, then under the condition (\ref{bbu}) it becomes $V=V'=0$, and the motion equation (\ref{motion2}) becomes $b^rR_{rr}=0$.
 In a  four dimensional spacetime, the gravitational equations (\ref{tt},\ref{rr},\ref{theta}) become,}
\begin{eqnarray}\label{}
&&(1+\ell)(2r\psi'-1)+e^{2\psi}(1-\Lambda r^2)=0\label{tt4}\\
&&\ell r^2(\phi''+\phi'^2-\phi'\psi')-2\ell r(\psi'+\phi')-2r\phi'+e^{2\psi}(1-\Lambda r^2)=0\label{rr4}\\
&&(1+\ell)[r^2(\phi''+\phi'^2-\phi'\psi')+r(\phi'-\psi')]+e^{2\psi}\Lambda r^2=0\label{th4},
\end{eqnarray}
and $R_{rr}=0$ becomes
\begin{eqnarray}\label{Rrr}
&&R_{rr}=\frac{2}{r}\psi'-(\phi''+\phi'^2-\phi'\psi')=0.
\end{eqnarray}
by using Eq. (\ref{Rrr}), the Eqs. (\ref{rr4}) and (\ref{th4}) become
\begin{eqnarray}\label{}
&&-(1+\ell)(2r\phi'+1)+e^{2\psi}(1-\Lambda r^2)=0\label{rrr4}\\
&&(1+\ell)r(\phi'+\psi')+e^{2\psi}\Lambda r^2=0\label{the4}.
\end{eqnarray}
Combining Eq. (\ref{rrr4}) and (\ref{tt4}), one can obtain $\phi'+\psi'=0$, which will lead Eq. (\ref{the4}) to the result $e^{2\psi}\Lambda r^2=0$, i.e., $\Lambda=0$. Therefore, it is proved that there is no black hole solution with nonzero $\Lambda$ for the above gravitational equations (\ref{tt4}) to (\ref{th4}) under the smooth potential $V=\kappa x^2/2$.

\section{An equivalent way to the first law}
 By using the thermodynamical volume $V$ (\ref{volume}) and the entropy $S$ (\ref{entropy}), one can rewrite the mass formula (\ref{hmass}) as
\begin{eqnarray}\label{}
M=M(S,P)=\frac{(D-2)S(r_h)}{4\pi \sqrt{1+\ell}r_h}+PV(r_h).
\end{eqnarray}
Then its total differential is
\begin{eqnarray}\label{fir}
dM=\Big(\frac{\partial M}{\partial S}\Big)_PdS+\Big(\frac{\partial M}{\partial P}\Big)_SdP.
\end{eqnarray}
In it, the partial derivative $(\frac{\partial M}{\partial P})_S=V$,
another partial derivative is
\begin{eqnarray}\label{first}
\Big(\frac{\partial M}{\partial S}\Big)_P=\Big(\frac{\partial M}{\partial S}\Big)_{P,V,r_h}+\Big(\frac{\partial M}{\partial r_h}\Big)_{P,S}\Big(\frac{dr_h}{dS}\Big),
\end{eqnarray}
whose result is the same as the Eq. (\ref{tem}).
Lastly Eq.(\ref{fir}) gives the first law of black hole thermodynamics,
\begin{eqnarray}\label{first}
dM=TdS+VdP.
\end{eqnarray}

\end{document}